\documentclass[aps,prl,twocolumn,superscriptaddress,floatfix]{revtex4}
\usepackage{graphicx}

\begin{document} 

\title{Search for the Rare Decay $K_{L}\rightarrow\pi^{0}\pi^{0}\gamma$}


\author{
E.~Abouzaid$^{4}$,
M.~Arenton$^{11}$,
A.R.~Barker$^{5,*}$,
L.~Bellantoni$^7$,
E.~Blucher$^4$,
G.J.~Bock$^7$,
E.~Cheu$^1$,
R.~Coleman$^7$,
M.D.~Corcoran$^{9}$,
G.~Corti$^{11}$,
B.~Cox$^{11}$,
A.R.~Erwin$^{12}$,
C.O.~Escobar$^{3}$,
A.~Glazov$^4$,
A.~Golossanov$^{11,7}$,
R.A.~Gomes$^{3}$,
P.~Gouffon$^{10}$,
Y.B.~Hsiung$^7$,
D.A.~Jensen$^7$,
R.~Kessler$^4$,
K.~Kotera$^8$,
A.~Ledovskoy$^{11}$,
P.L.~McBride$^7$,
E.~Monnier$^{4,**}$,
H.~Nguyen$^7$,
R.~Niclasen$^{5}$,
D.G.~Phillips~II$^{11}$,
H.~Ping$^{12}$,
E.J.~Ramberg$^7$,
R.E.~Ray$^7$,
M. Ronquest$^{11}$,
E.~Santos$^{10}$,
J.~Shields$^{11}$,
W.~Slater$^2$,
D.~Smith$^{11,\dagger}$,
N.~Solomey$^4$,
E.C.~Swallow$^{4,6}$,
P.A.~Toale$^5$,
R.~Tschirhart$^7$,
C.~Velissaris$^{12}$,
Y.W.~Wah$^4$,
J.~Wang$^1$,
H.B.~White$^7$,
J.~Whitmore$^7$,
M.~Wilking$^5$,
R.~Winston$^4$,
E.T.~Worcester$^4$,
M.~Worcester$^4$,
T.~Yamanaka$^8$,
E.D.~Zimmerman$^5$,
R.F.~Zukanovich$^{10}$\\
(KTeV Collaboration)} 

\affiliation{
$^{1}$University of Arizona, Tucson, Arizona 85721 \\
$^{2}$University of California at Los Angeles, Los Angeles, California 90095\\
$^{3}$Universidade Estadual de Campinas, Campinas, Brazil 13083-970\\
$^{4}$The Enrico Fermi Institute, The University of Chicago, Chicago, Illinois
60637\\
$^{5}$University of Colorado, Boulder Colorado 80309\\
$^{6}$Elmhurst College, Elmhurst, Illinois 60126\\
$^{7}$Fermi National Accelerator Laboratory, Batavia, Illinois 60510\\
$^{8}$Osaka University, Toyonaka, Osaka 560-0043 Japan\\
$^{9}$Rice University, Houston, Texas 77005\\
$^{10}$Universidade de Sao Paulo, Sao Paulo, Brazil 05315-970\\
$^{11}$University of Virginia, Charlottesville, Virginia 22901\\
$^{12}$University of Wisconsin, Madison, Wisconsin 53706\\ }

\begin{abstract} 
The KTeV E799 experiment has conducted a search for the rare
decay $K_{L}\rightarrow\pi^{0}\pi^{0}\gamma$ via the topology 
$K_{L}\rightarrow\pi^{0}\pi^{0}_D\gamma$ (where
$\pi^0_D\rightarrow\gamma e^+e^-$).  
Due to Bose statistics of the $\pi^0$ pair and the real nature of
the photon, the $K_{L}\rightarrow\pi^{0}\pi^{0}\gamma$ decay is restricted to
proceed at lowest order by the CP conserving direct emission (DE) of an E2 
electric quadrupole photon.  The rate of this decay is interesting theoretically since chiral 
perturbation theory predicts that this process vanishes at level $O(p^4)$.  
Therefore, this mode probes chiral perturbation theory at $O(p^6)$.
In this paper we report a determination of an upper limit of 
$2.43\times 10^{-7}$  (90\% CL) for $K_{L}\rightarrow\pi^{0}\pi^{0}\gamma$.
This is approximately a factor of 20 lower than previous results.
\\ \\
\vspace{0.25in}
\noindent
PACS numbers: 13.20.Eb, 13.25.Es
\end{abstract}

\maketitle

\section {I. INTRODUCTION}
While measurements of branching ratios for kaon decay modes such as 
$K_L\rightarrow\gamma\gamma$ have shown good agreement~\cite{ref:0.5} 
with chiral perturbation theory (ChPT) calculations to order $O(p^4)$, 
ChPT calculation to order $O(p^6)$ have been difficult to check due
to the presence of the large lower order terms. Since the 
$K_L\rightarrow\pi^0\pi^0\gamma$ mode is forbidden to order $O(p^4)$
in ChPT~\cite{ref:0}, this mode presents 
an opportunity to check ChPT at higher orders. Therefore an experimental  
measurement of this mode's branching ratio is of interest.  At present, 
a chiral dimensional analysis~\cite{ref:0.75} 
for the $K_L\rightarrow\pi^0\pi^0\gamma$ mode results in a branching ratio 
of $7\times10^{-11}$.  Because of the Bose statistics of the $\pi^0\pi^0$ pair and the 
presence of a real direct emission $\gamma$, the
$K_{L}\rightarrow\pi^{0}\pi^{0}\gamma$ decay proceeds at lowest order 
via the CP conserving direct emission of a electric quadrupole (E2) photon. 
This causes a large suppression of this mode relative to the associated 
$K_L\rightarrow\pi^+\pi^-\gamma$ mode.  Accordingly, observation of 
$K_L\rightarrow\pi^0\pi^0\gamma$ would give a measurement of E2 quadrupole
emission which is difficult to extract from the E1 and M1 dominated
$K_L\rightarrow\pi^+\pi^-\gamma$ mode. The M1 amplitude determined from
$K_L\rightarrow\pi^+\pi^-\gamma$ decay together with a plausible estimate 
for the E2 amplitude~\cite{ref:1}, given by
\begin{equation}
\nonumber A(\pi^0\pi^0\gamma)=\frac{g_{E2}}{M^4_K}\frac{(p_1-p_2)\cdot k}{\Lambda^2}
\times \epsilon\cdot (p_1 k\cdot p_2- p_2 k\cdot p_1),
\end{equation}
\noindent can be used to obtain an estimate of the  branching
ratio for $K_L\rightarrow\pi^0\pi^0\gamma$. Here $p_1$ and $p_2$ are the momenta of the pions and k
and $\epsilon$ are the momentum and polarization of the photon respectively. 
$g_{E2}$ is the coupling of the E2 amplitude. $\Lambda^{-1}$, a mass parameter that measures
the extent of the region over which the interaction takes place, is assumed
to be of order of the $\rho$ meson mass. The CP violating direct emission of
a M2 magnetic quadrupole photon is much smaller than the E2 transition and can be ignored. 
According to Ref.~\cite{ref:1}, the decay rate for $K_{L}\rightarrow\pi^{0}\pi^{0}\gamma$ can be 
estimated by comparing it to the M1 direct emission (DE) amplitude 
of the $K_L\rightarrow\pi^+\pi^-\gamma$ decay
\begin{equation}
A(\pi^+\pi^-\gamma)=\frac{g_{M1}}{M^4_K}\epsilon_{\mu\nu\rho\sigma}\epsilon^{\mu}k^{\nu}p^{\rho}_{+}p^{\sigma}_{-}
\end{equation}
\noindent Employing this amplitude, the ratio
\begin{equation}
\frac{\Gamma(K_L\rightarrow\pi^0\pi^0\gamma)}{\Gamma(K_L\rightarrow\pi^+\pi^-\gamma)_{DE}}=
\frac{1}{2}(\frac{g_{E2}}{g_{M1}})^2(\frac{M_K^2}{\Lambda^2})^2(4\times10^{-3}).
\end{equation}
\noindent can be formed. Using the fraction 
of $\Gamma(K_L\rightarrow\pi^+\pi^-\gamma)$ to $\Gamma(K_L\rightarrow\pi^+\pi^-)$
due to M1 direct emission of $(14.2\pm0.28)\times 10^{-3}$ from Ref.~\cite{ref:2} and a
$K_L\rightarrow\pi^+\pi^-$ branching ratio of $(2.090\pm0.025)\times 10^{-3}$
from the PDG~\cite{ref:3}, a branching ratio for M1 DE part of the
$K_L\rightarrow\pi^+\pi^-\gamma$ 
decay of $2.96\times10^{-5}$ ($E_{\gamma}\geq 20 $MeV) is obtained. Using Eq. 3
above, this gives an estimated branching ratio of $1.08\times 10^{-8}$ for
$K_L\rightarrow\pi^0\pi^0\gamma$ (under the assumption that $g_{E2}$ has
similar magnitude to $g_{M1}$).  

The KTeV collaboration previously reported an upper limit of 
of $5.4\times10^{-9}$ (90\% CL)~\cite{ref:4} for the 
related decay $K_{L}\rightarrow\pi^{0}\pi^{0}e^+e^-$ in which the direct 
emission E2 photon was virtual. However, we point out that this decay, 
in contrast to the $K_{L}\rightarrow\pi^{0}\pi^{0}\gamma$
decay where the photon is real, can proceed via an additional process
forbidden for a real photon. The virtual photon is emitted in a J=0 state
from the $K_L$ allowing a transition of the $K_L$ to a $K_S$ followed by the CP
allowed decay $K_S\rightarrow\pi^0\pi^0$.  This is the so-called charge radius 
amplitude~\cite{ref:5}. The rates for these two decays would have
similar branching ratios except for the the charge radius amplitude.

The previous best upper limit of $5\times 10^{-6}$
(90\% CL) for the $K_{L}\rightarrow\pi^{0}\pi^{0}\gamma$ decay 
obtained by the NA31 experiment~\cite{ref:7} was achieved 
by searching for this mode in events with five photons 
in the final state. Due to the particular configuration 
of triggers and prescale factors in the
KTeV experiment, the most sensitive method was to require 
that one of the neutral pions undergo Dalitz decay $\pi^0\rightarrow e^+e^-\gamma$.

\section{II. The KTeV E799 Experiment}
The search for the $K_{L}\rightarrow\pi^0\pi^0_D\gamma$ mode was performed 
using the 1997 and 1999 runs of KTeV E799 II at Fermi National
Accelerator Laboratory. The KTeV E799 experiment used two almost
parallel $K_L$ beam lines generated by interactions of 800 GeV/c protons in
a BeO target some 90 meters upstream of the KTeV spectrometer.  The KTeV
spectrometer consisted of a 70 meter evacuated decay tube followed by, in
sequence, two stations of drift chambers, a large aperture dipole magnet, two
more stations of drift chambers, a multi-plane transition radiation detector,
a 3100 element CsI electromagnetic calorimeter and a muon detector.

The 1997 and 1999 runs differed in the following ways.  The spill length was
doubled from 20 in the 1997 run to 40 seconds in the 1999 run.  The proton
intensity on the BeO neutral kaon production target was increased from 
$4\times10^{12}$ in the 1997 run to $6-10\times10^{12}$ per spill in the 1999 run.  
Another important difference between the 1997 and 1999 run was that the
magnetic field was decreased from a transverse deflection  of 205 MeV/c in
1997 to 150 MeV/c in 1999 to increase acceptance for some of the neutral kaon
modes.  The trigger which demanded two charged tracks and at least four
electromagnetic clusters was loosened midway during the 1997 run by changing
the minimum thresholds for energy in the electromagnetic calorimeter and 
remained loose for the 1999 run. 

\subsection{II A. The $K_L\rightarrow\pi^0\pi^0_D\gamma$ Signal Criteria}
The $K_L\rightarrow\pi^0\pi^0_D\gamma$ final state consists of four photons
plus an $e^+e^-$ pair.  The KTeV analysis required six electromagnetic 
showers ($E\geq$ 0.6 GeV) 
in the CsI calorimeter, two of which were associated with two reconstructed
charged tracks. The two charged tracks were required to be consistent with the
two track trigger, form a good charged track vertex, have opposite charge, and
have $0.95\leq E/p \leq 1.05$ (where $E$ is the energy of the associated shower in the CsI
calorimeter and $p$ is the momentum determined from magnetic bending). 

Events passing these cuts were required to satisfy
several additional cuts to select $K_{L}\rightarrow\pi^{0}\pi^{0}\gamma$ and
to reject the major background due to $K_L\rightarrow\pi^0\pi^0\pi^0_D$
decays.  These events can contribute to the background if one of the photons
was not detected or two of the photons reconstructed as single photon.
Additional cuts were performed including a vertex cut 
determined by cycling over all $\gamma\gamma$ and $e^+e^-\gamma$
combinations in a given event for all possible vertex positions
to determine the vertex where the $M_{e^+e^-\gamma}$
and $M_{\gamma\gamma}$ were closest to the $\pi^0$ mass.  The 
vertex thus obtained was required
to be between 95 and 150 meters from the BeO target in the KTeV decay
volume and to have a good vertex $\chi^2$.
Another cut was made on the variable ($P_L^2)_{\pi^0}$ defined as 
the longitudinal momentum squared of the $\pi^0$ in the frame in which the 
momentum of the $\pi^0\pi^0_D$ pair is totally transverse.  
If the event is a $K_L\rightarrow\pi^0\pi^0\pi^0_D$ decay,
($P_L^2)_{\pi^0}$ is greater than zero.  If the event is a 
$K_{L}\rightarrow\pi^0\pi^0_D\gamma$ decay, ($P_L^2)_{\pi^0}$ 
is less than zero (modulo resolution). Accordingly ($P_L^2)_{\pi^0}$ was 
required $\leq -0.005$ GeV$^2$/c$^2$ in order to select
$K_L\rightarrow\pi^0\pi^0_D\gamma$ decays and reject $K_L\rightarrow\pi^0\pi^0\pi^0_D$.

The analysis cuts were varied slightly between the 1997 and 1999 run.  The
main changes were to alter the shape of the  $p^2_t(ee\gamma\gamma\gamma\gamma)$ and 
$M_{ee\gamma\gamma\gamma\gamma}$ signal region and to tighten 
the decay vertex position and the overlapping shower cuts to reject 
more background.

\subsection{II B. Backgrounds}
Monte Carlos were developed for the signal $K_{L}\rightarrow\pi^0\pi^0_D\gamma$
mode incorporating the model of Ref.~\cite{ref:1} as well as for 
the various backgrounds, and the normalization mode incorporating 
all known features of the E799 spectrometer and beam. The 
various $K_L$ modes were generated using experimentally
determined decay parameters.  Trigger and analysis cuts applied to 
the Monte Carlos were the same as those applied to the data.  Background 
and normalization mode Monte Carlos were generated to produce Monte
Carlo samples several times the 1997 and 1999 run $K_L$ fluxes.

\begin{figure}
\begin{center}
\includegraphics[width=0.40\textwidth]{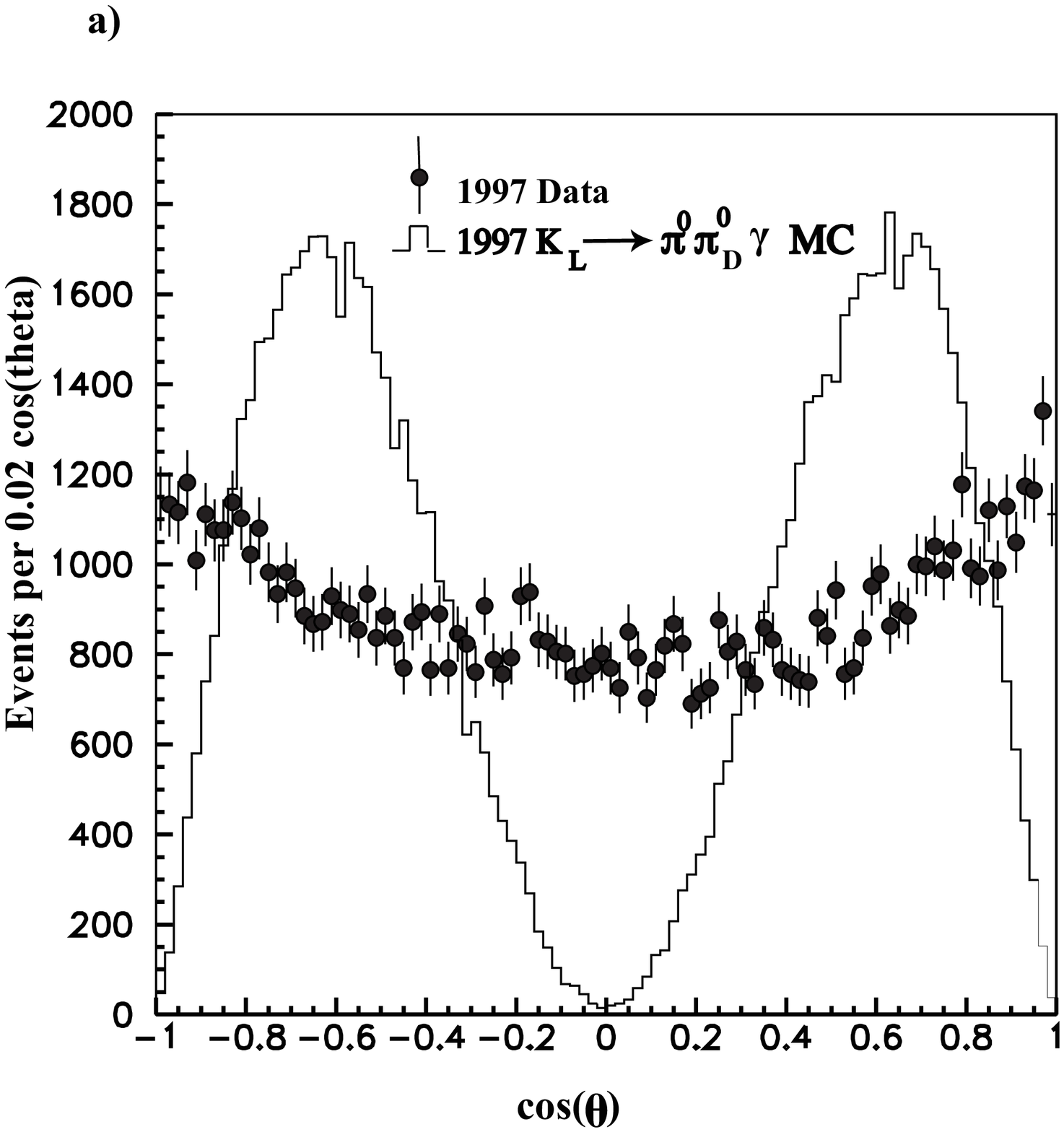}
\includegraphics[width=0.40\textwidth]{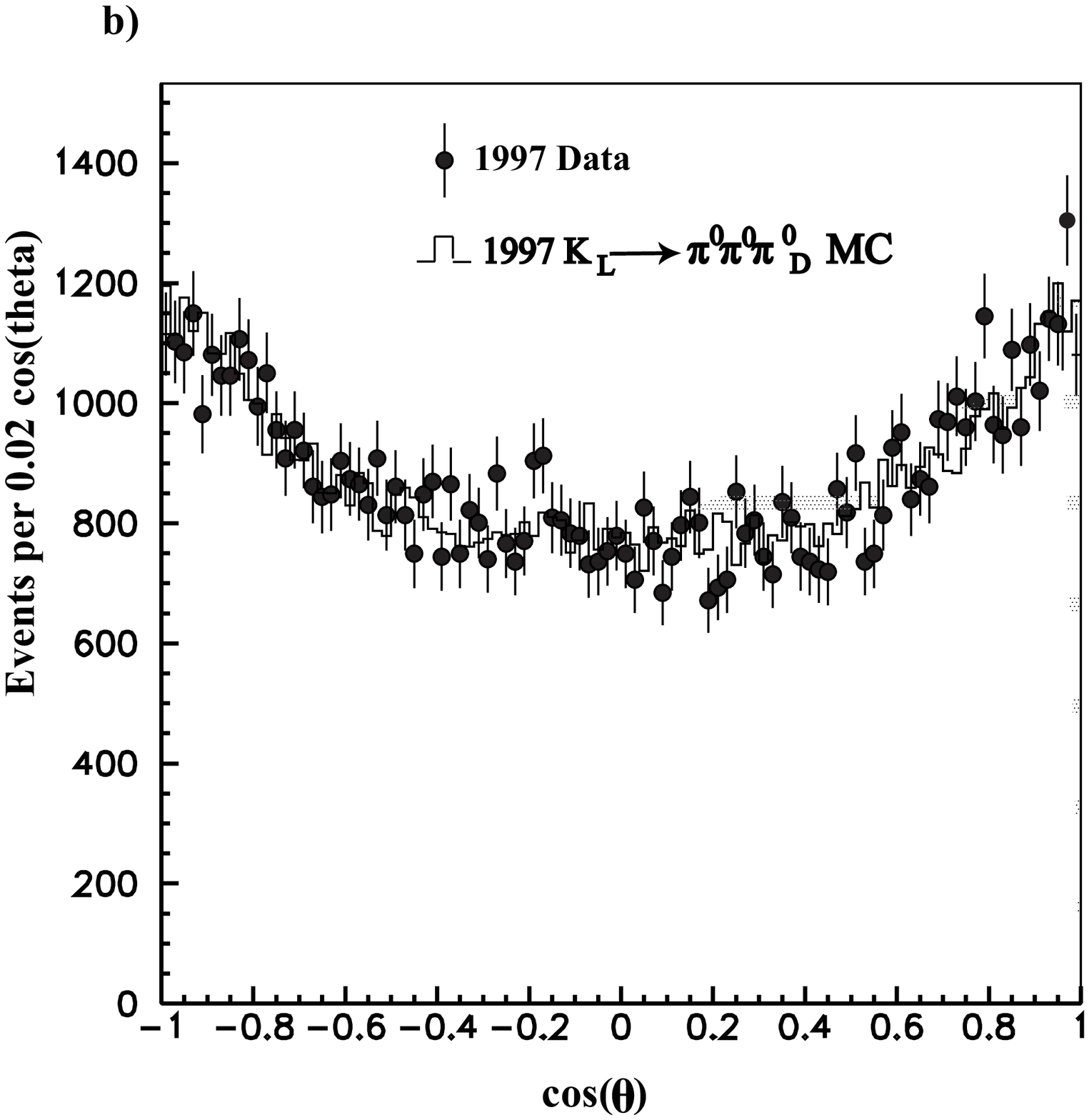}
\end{center}
\caption{a) comparison of cos($\theta$) for a $K_{L}\rightarrow\pi^0\pi^0_D\gamma$ signal MC vs.
cos($\theta$) as observed in the 1997 data; b) comparison of cos($\theta$) for
the $K_{L}\rightarrow\pi^0\pi^0\pi^0_D$ background MC vs cos($\theta$)
for the 1997 data}
\label{Fig:1.25}
\end{figure}

As indicated in Fig.~\ref{Fig:1.25}, a comparisons of a $K_{L}\rightarrow\pi^0\pi^0_D\gamma$
and a background Monte Carlo $K_{L}\rightarrow\pi^0\pi^0\pi^0_D$ with
the 1997 data shows that the angle $\theta$ between 
the direct emission photon and one of the two $\pi^0$'s in the $\pi^0\pi^0$ 
center of mass is peaked for the signal mode at cos$\theta = \pm 0.6$ and flat 
for $K_L\rightarrow\pi^0\pi^0\pi^0_D$. The same is true for the 1999 data.

\begin{figure}
\begin{center}
\includegraphics[width=0.50\textwidth]{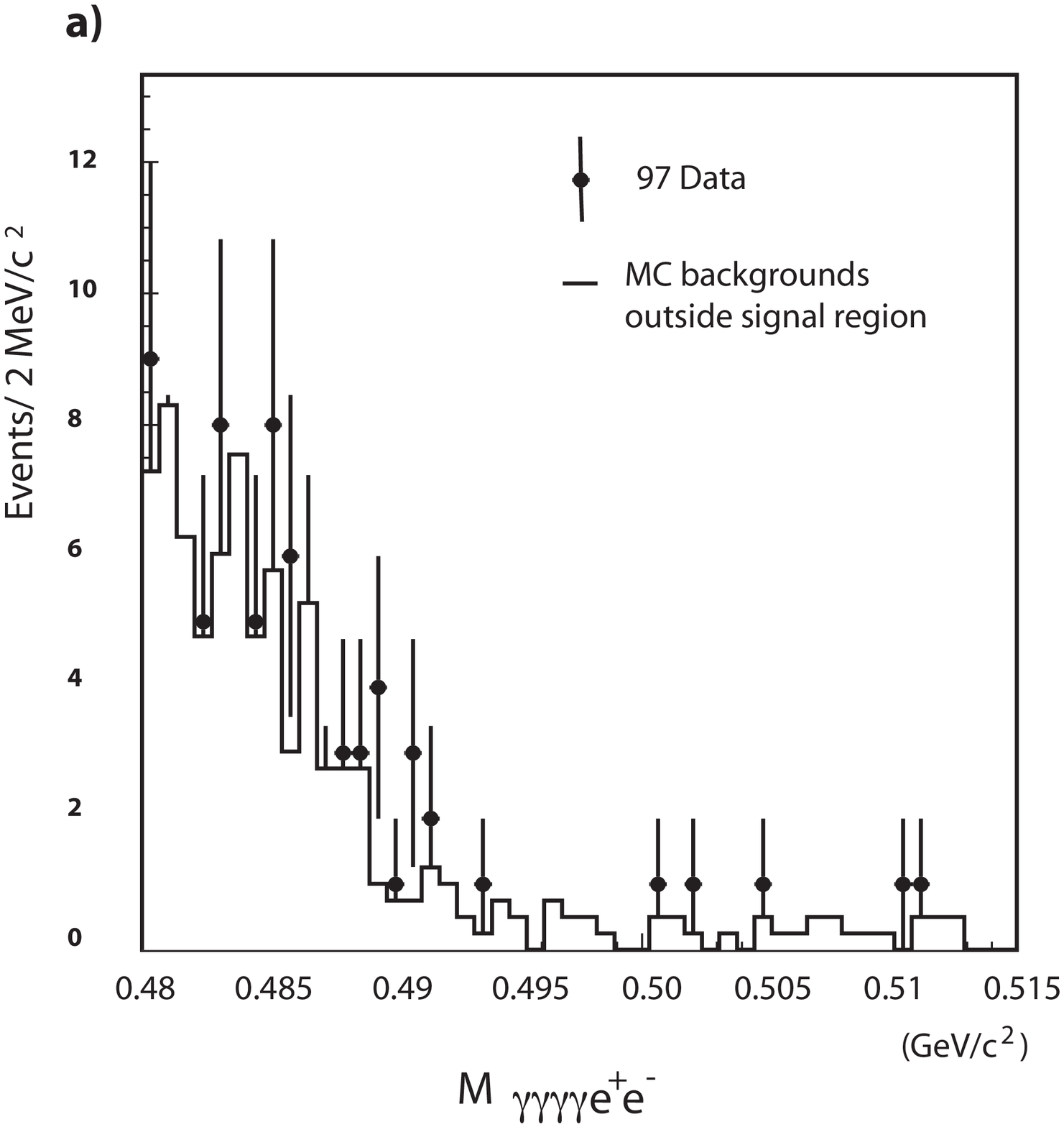}
\includegraphics[width=0.50\textwidth]{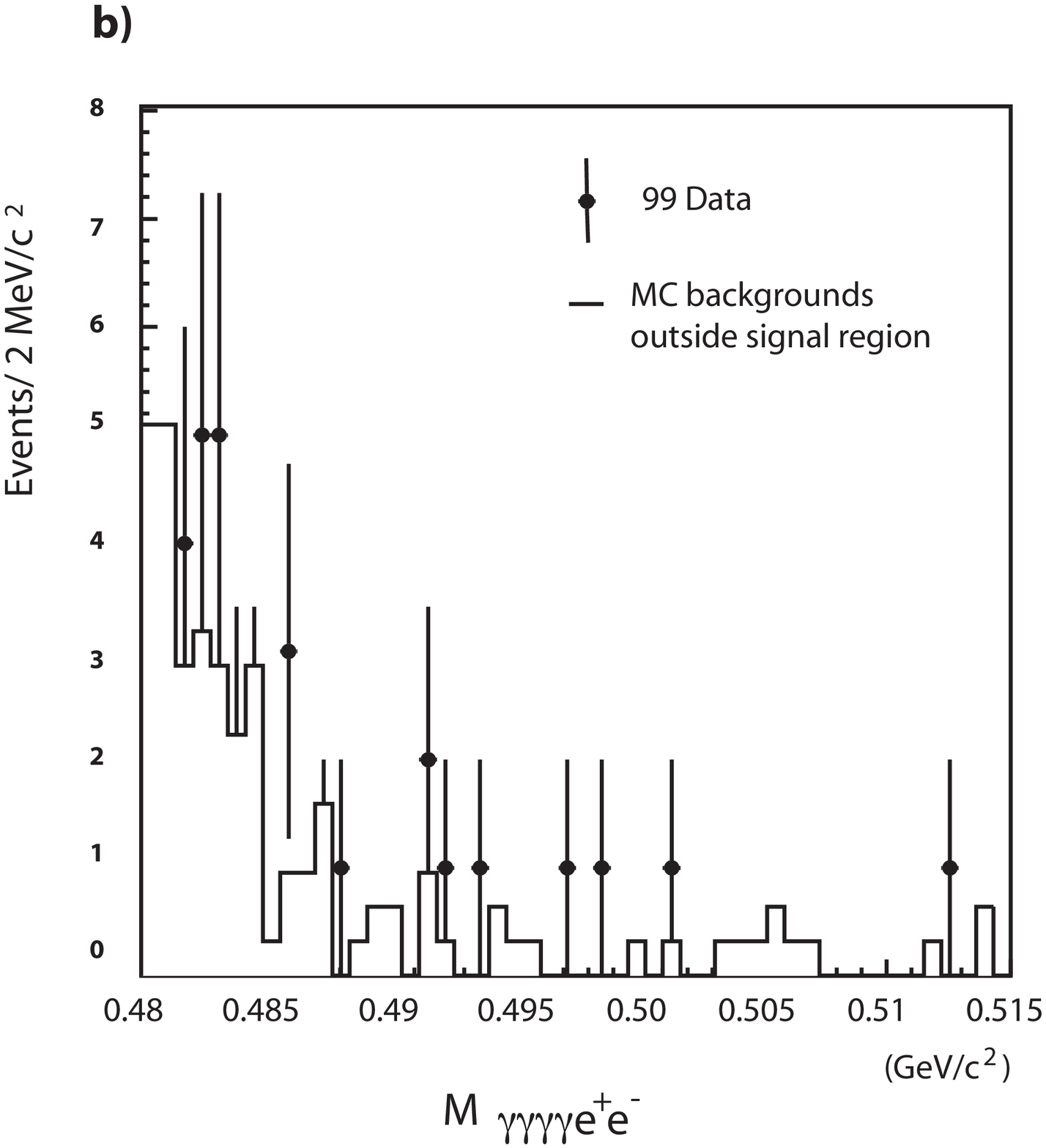}
\end{center}
\caption{Data and Monte Carlo background comparisons passing all cuts except
the signal region contours of Fig.~\ref{Fig:2} for a) the 1997 run; b) the
1999 run.}
\label{Fig:1.5}
\end{figure}

Therefore a final cut to 
select $K_{L}\rightarrow\pi^0\pi^0_D\gamma$ was made requiring events 
to have $0.25 \leq |$cos$\theta|\leq 0.9$ to improve signal to background.
Several other decays were investigated to see if there was any
contribution to background after the cuts to select $K_{L}\rightarrow\pi^0\pi^0_D\gamma$.
These modes included
$K_L\rightarrow\pi^0\pi^0_D\pi^0_D$ where a Dalitz pair is undetected,
$K_L\rightarrow\pi^0\pi^0\pi^0$ followed by the decay of a $\pi^0$ into an
$e^+e^-$ pair, $K_L\rightarrow\pi^0\pi^0\pi^0$ followed by the decay of a 
$\pi^0$ into $e^+e^-e^+e^-$ with one of the $e^+e^-$ pairs undetected, 
$K_L\rightarrow\pi^0e^+e^-\gamma$ decays with accidental photons, and
$K_L\rightarrow\pi^0\pi^0_D$ plus accidental photons.  In no case did
any of these modes contribute significantly to the background. 
The agreement of the Monte Carlo simulation of the remaining
background with the 1997 and 1999 data 
outside the signal region is shown in Fig.~\ref{Fig:1.5}.

\section{III. Search for a $K_L\rightarrow\pi^0\pi^0_D\gamma$ Signal}
The signal regions for the 1997 and 1999 data were based on the 
$M_{ee\gamma\gamma\gamma\gamma}$ and $p^2_t(ee\gamma\gamma\gamma\gamma)$ 
resolutions calculated using a signal mode Monte Carlo.  Here $p^2_t$ is
measured relative to the direction of the $K_L$ determined by the line
connecting the BeO target center and the decay vertex. For the 1997 data, 
the signal region was chosen to be rectangular with 0.494 GeV/c$^2\leq$ M
$\leq$ 0.501 GeV/c$^2$ and $p_t^2\leq$ 0.00015 (GeV/c)$^2$.  In the later 
analysis of the 1999 data, it was decided to use a contour containing 68\% 
of the signal Monte Carlo events as determined from a joint probability distribution
based on the signal Monte Carlo resolutions
for $p^2_t(ee\gamma\gamma\gamma\gamma)$ and $M_{ee\gamma\gamma\gamma\gamma}$. 
Since the signal box had already been opened for the 1997 data (and no events were 
observed as discussed below), the rectangular 
signal region was kept for the 97 data. 

\begin{figure}
\begin{center}
\includegraphics[width=0.32\textwidth]{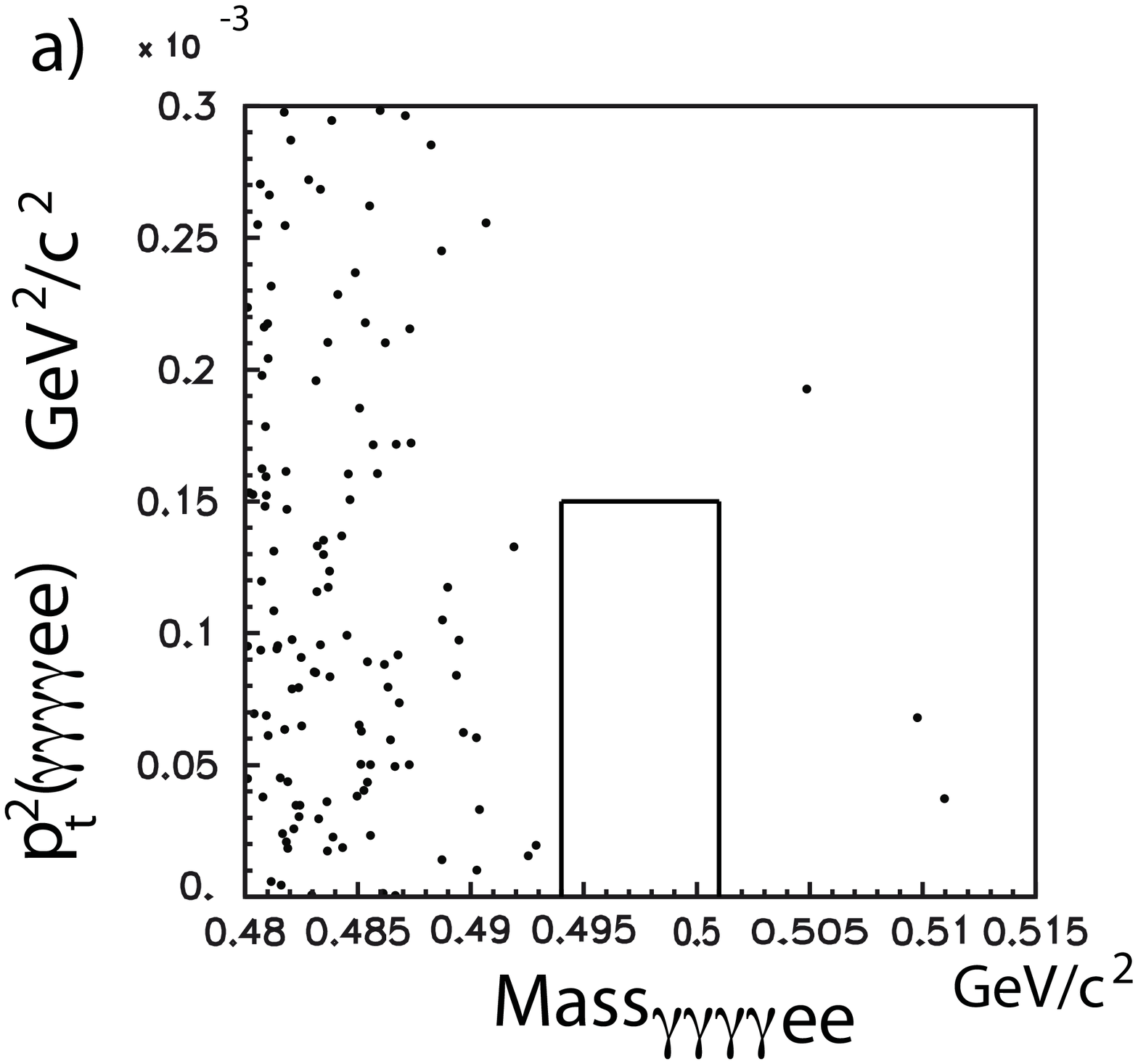}
\includegraphics[width=0.32\textwidth]{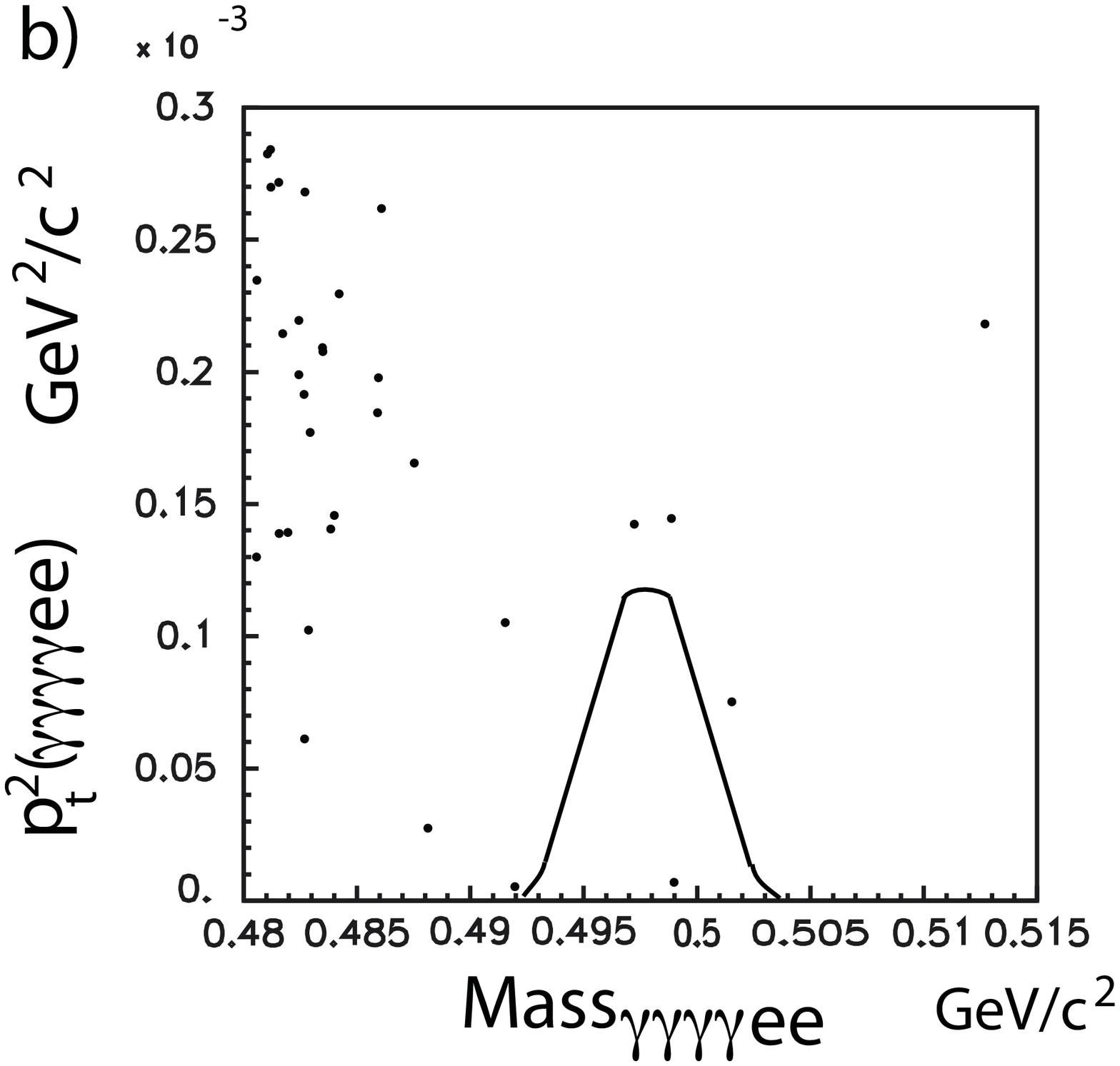}
\end{center}
\caption{a) $M_{ee\gamma\gamma\gamma\gamma}$
vs. $p^2_t(ee\gamma\gamma\gamma\gamma)$ for the 1997 KTeV data;
b) $M_{ee\gamma\gamma\gamma\gamma}$
vs. $p^2_t(ee\gamma\gamma\gamma\gamma)$ for the 1999 KTeV data}
\label{Fig:2}
\end{figure}

As shown in Fig.~\ref{Fig:2}a, no events were observed in the 1997 data in this
region when the box was opened after the blind analysis was completed.
From the  $K_L\rightarrow\pi^0\pi^0\pi^0_D$ background Monte Carlo, 
we expected $0.83\pm0.41$ background events in the 1997 run.  As shown in
Fig.~\ref{Fig:2}b,  one data event was found in the 1999 
signal region while the background $K_L\rightarrow\pi^0\pi^0\pi^0_D$ Monte Carlo 
yielded no events in the signal region for four 
equivalent $K_L$ fluxes. These Monte Carlo background estimates for the
1997 and 1999 runs are consistent with the background estimates based on
mass and $p^2_t$ sideband projection from the data.  
   
\section {IV. Determination of the Upper Limit for $K_L\rightarrow\pi^0\pi^0_D\gamma$} 
Since the $K_L\rightarrow\pi^0\pi^0\pi^0_D$ branching ratio is well measured 
and since events where a photon was lost down the beam hole had the same
topology as the signal and were copious, these events were used for the normalization 
for the experiment. The reason for selecting this topology was to minimize
the systematics due to trigger and event configuration between the signal and
normalization modes.

The criteria to select these normalization mode events 
included the same topology criteria, the vertex position cut, and the $\pi^0$
mass cuts as the signal mode.  Cuts that
were special to the selection of the normalization mode included 
a requirement that $(P_L^2)_{\pi^0}\geq 0.0$.  In addition, to insure that
the event was consistent with a $K_L\rightarrow\pi^0\pi^0\pi^0_D$ decay with a photon
going down the beam hole, the missing photon was reconstructed by
assuming a kaon mass and net zero $p_t^2$ for the event. This resulted in
two solutions for energy and trajectory of the missing photon. If one of the solutions
for the missing photon combined with the unpaired photon in the observed event to
form a $\pi^0$ mass within 0.05 GeV/c$^2$, if the other 
three photons and the $e^+e^-$ in the event paired
to form $\pi^0$'s, and if the missing photon's calculated trajectory 
pointed to either beam hole, the event was included in the normalization sample.
Normalization mode yields of 7725 and 9024 $K_L\rightarrow\pi^0\pi^0\pi^0_D$ 
were obtained for the 1997 and 1999 data using these criteria.

\subsection{IV A. The Single Event Sensitivity for $K_L\rightarrow\pi^0\pi^0_D\gamma$ Signal}
We can avoid the extra error due to the branching ratio uncertainty 
in the Dalitz decay by noting that the
$K_L\rightarrow \pi^0\pi^0_D\gamma$ single event sensitivities ({\em SES}) are given by
\begin{equation}
SES(\pi^0\pi^0_D\gamma)_{97,99}=\frac{1}{A(\pi^0\pi^0_D\gamma)_{97,99}N(K_L)_{97,99}}
\end{equation}
\noindent where A$(\pi^0\pi^0_D\gamma)_{97,99}$ are the acceptances for
the $K_L\rightarrow\pi^0\pi^0_D\gamma$ Dalitz decay mode and N$(K_L)_{97,99}$ are
the $K_L$ fluxes for the 1997 and 1999 modes.  The fluxes are
\begin{equation}
N(K_L)_{97,99}=\frac{N(3\pi^0_D)^{norm}_{97,99}}
{A(3\pi^0_D)^{norm}_{97,99}BR(K_L\rightarrow\pi^0\pi^0\pi^0_D)}
\end{equation}
\noindent so the {\em SES}'s for
$K_L\rightarrow\pi^0\pi^0\gamma\rightarrow\gamma\gamma\gamma\gamma\gamma$ are
given by
\begin{equation}
SES_{97,99}=\frac{3BR(K_L\rightarrow\pi^0\pi^0\pi^0)BR(\pi^0\rightarrow\gamma\gamma)A(3\pi^0_D)^{norm}_{97.99}}
{2A(\pi^0\pi^0_D\gamma)_{97,99}N(3\pi^0_D)^{norm}_{97,99}}
\end{equation}
\noindent where $N(3\pi^0_D)^{norm}_{97,99}$ and $A(3\pi^0_D)^{norm}_{97,99}$ are
the number and the acceptance of normalization mode
$K_L\rightarrow\pi^0\pi^0\pi^0_D$ events 
in the 97 and 99 runs.  The factor of 2 and 3 arise
from the number of $\pi^0$'s that can produce a Dalitz pair in the
signal and normalization modes.  Note that the Dalitz decay branching ratio 
cancels out in the {\em SES} for $K_L\rightarrow\pi^0\pi^0\gamma$.  

From the $K_L\rightarrow\pi^0\pi^0_D\gamma$ and 
$K_L\rightarrow\pi^0\pi^0\pi^0_D$ Monte Carlos, the signal and normalization 
acceptances were determined to be 0.101\% and $4.21\times 10^{-6}$ for the 1997 run and
0.085\% and $3.26\times 10^{-6}$ for the 1999 runs 
respectively. Using Eq. 6 and the measured 19.56$\pm$0.12\% branching ratio 
for $K_L\rightarrow\pi^0\pi^0\pi^0$~\cite{ref:8.5}, the 1997 and 1999 {\em SES}'s were
determined to be $1.56\times 10^{-7}$ and $1.25\times 10^{-7}$ respectively.
In addition, $K_L$ fluxes of 2.68$\times 10^{11}$ and 3.99$\times 10^{11}$ 
were obtained for the 1997 and 1999 runs. The {\em SES}'s for 97/99 were combined using
\begin{equation}
\frac{1}{SES_{combined}}=\frac{1}{SES_{97}}+\frac{1}{SES_{99}}
\end{equation}
\noindent resulting in a combined $SES$ of $6.93\times 10^{-8}$. 

\subsection {IV B. Systematic Errors}
Systematic errors for the upper limit could arise from several sources.  Among them
was the error in the branching ratio for $K_L\rightarrow\pi^0\pi^0\pi^0$. There
were also disagreements between the distributions of the $K_L\rightarrow\pi^0\pi^0\pi^0_D$
normalization mode data and Monte Carlo. The effect of these
disagreements on the {\em SES} were studied by adjusting the Monte Carlos to
eliminate the disagreements and seeing
what changes in the {\em SES} took place.  There could also be  disagreements 
between the $K_{L}\rightarrow\pi^0\pi^0_D\gamma$ data and its
Monte Carlo.  A disagreement in signal mode data and Monte Carlo could not
be checked directly since no signal is observed. However, from an
inspection of the calculation of the {\em SES}, it can be seen that the {\em SES} is proportional to
$(N_{signal}/A_{signal})/(N_{norm}/A_{norm})$.  So, to the level that the
normalization and signal mode topologies are similar, it was expected that differences
in acceptance between the Monte Carlo and the data would tend to cancel.
Studies were done of the effect of disagreements between the signal mode and
normalization mode Monte Carlos by adjusting the Monte Carlos until they
agreed and determining the effect on the result.

\begin{table} 
\begin{tabular}{|l|c|}
\hline
Error on the 3$\pi^0$ branching ratio            &  0.61\%                \\
Error on the $\pi^0\rightarrow\gamma\gamma$ branching ratio & 3.24\%      \\
Normalization Monte Carlo/data disagreements              &  3.57\%                \\
Signal/normalization Monte Carlo disagreements    &  5.35\%                \\
\hline
Total Systematic Error                           &  7.23\%                \\
\hline
\end{tabular}
   \caption{Systematic uncertainties for $K_L\rightarrow\pi^0\pi^0\gamma$ {\em SES}}
 \label{systematics}
\end{table}
The percentage systematic errors in the {\em SES} due to these sources are shown in
Table~\ref{systematics}. Adding these systematic errors 
in Table I in quadrature, we obtain a total systematic error in the {\em SES} of
7.23\%.  Taking into account the statistical error of 1.09\%, this leads to a
total error of 7.31\% resulting in a {\em SES} error of $5.07\times10^{-9}$. 

\subsection {IV C. Calculation of the Upper Limit for $K_L\rightarrow\pi^0\pi^0_D\gamma$ Signal}
The method of Ref.~\cite{ref:10} was used to obtain an upper limit 
for $K_L\rightarrow\pi^0\pi^0\gamma$.  If $n_{exp}$ is the expected number 
of signal plus background events in the signal box and $n_{bkg}$ is the 
expected number of background events, then the probability for observing 
$n$ events is given by a Poisson distribution $P(n_{exp},n)$ whose mean is
\begin{equation}
n_{exp}=n_{bkg}+BR(K_L\rightarrow\pi^0\pi^0\gamma))/SES
\end{equation}
\noindent The expected and observed backgrounds for the 97 and
99 runs were added to form the final observed and expected backgrounds.
A confidence region was constructed containing 90\% of the Poisson
distribution by varying the $K_L\rightarrow\pi^0\pi^0\gamma$ branching ratio
between 0 to $5\times10^{-7}$.  Using this confidence region, a 90\% CL upper limit for the
BR($K_L\rightarrow\pi^0\pi^0\gamma$) of $2.43\times 10^{-7}$ was determined, assuming
the E2 model of Ref.~\cite{ref:1}. The {\em SES} error was incorporated by varying
the {\em SES} over the 90\% CL range of the {\em SES} indicated by its combined statistical and systematic
error.

\section {V. Conclusions}
In conclusion,  the KTeV collaboration has obtained 
an upper limit for the branching ratio for the rare decay
$K_L\rightarrow\pi^0\pi^0_D\gamma$ of $2.43\times 10^{-7}$ 
using the 1997 and 1999 data and assuming that the decay
proceeds mainly via direct E2 photon emission. This limit is 
approximately twenty times lower than the best published upper
limit~\cite{ref:7}, providing a much more stringent upper limit 
for ChPT theoretical calculations of the E2 amplitude. 

\section {VI. Acknowledgments}
We thank the FNAL staff for their contributions.
This work was supported by the U.S. Department of Energy, the U.S. National Science Foundation, 
the Ministry of Education and Science of Japan, the Fundao de Amparo a
Pesquisa do Estado de So Paulo-FAPESP, the Conselho Nacional de
Desenvolvimento Cientifico e Tecnologico-CNPq, and the CAPES-Ministerio da Educao. 

\vspace{0.25cm}
\noindent $^{\dagger}$ Correspondence should be addressed to David Smith at
poldybloom@yahoo.com \\
\noindent $^{*}$ Deceased \\
\noindent $^{**}$ Permanent address C.P.P. Marseille/C.N.R.S., France\\

\end{document}